\documentclass[showpacs, amsmath, amssymb,superscriptaddress,twocolumn,nofootinbib,pra,aps]{revtex4-1}
\usepackage{color}
\usepackage[usenames,dvipsnames]{xcolor}
\usepackage{amsfonts,amssymb}
\usepackage{amsbsy}
\usepackage{enumitem}

\newcommand{\be}{\begin{eqnarray}}
\newcommand{\ee}{\end{eqnarray}}
\def\({\left(}
\def\){\right)}
\def\[{\left[}
\def\]{\right]}
\def\vect#1{\mbox{\boldmath $#1$}}

\newcommand{\braket}[1]{\left\langle #1 \right\rangle}
\newcommand{\bra}[1]{\left\langle #1 \right|}
\newcommand{\ket}[1]{\left| #1 \right\rangle}

\newcommand{\tr}{\mathrm{tr}}
\newcommand{\sla}[1]{\rlap{\kern .15em /}#1}

\newcommand{\eref}[1]{(\ref{#1})}

\newcommand{\eq}[1]{Eq. \eref{#1}}

\def\su{\mathfrak{su}}
\def\so{\mathfrak{so}}
\def \sig#1#2 {\sigma_{#1} \otimes \sigma_{#2}}
\newcommand{\ad}{\mathrm{ad}}

\begin{document}

\title{Dynamical Invariants for Quantum Control of Four--Level Systems}

\author{Utkan G\"ung\"ord\"u}
\email{utkan@alice.math.kindai.ac.jp}
\affiliation{Research Center for Quantum Computing, Interdisciplinary Graduate School of Science and Engineering, Kinki University, 3-4-1 Kowakae, Higashi-Osaka, Osaka 577-8502, Japan}

\author{Yidun Wan}
\email{ywan@alice.math.kindai.ac.jp}
\affiliation{Research Center for Quantum Computing, Interdisciplinary Graduate School of Science and Engineering, Kinki University, 3-4-1 Kowakae, Higashi-Osaka, Osaka 577-8502, Japan}

\author{Mohammad Ali Fasihi}
\email{fasihi@alice.math.kindai.ac.jp}
\affiliation{Research Center for Quantum Computing, Interdisciplinary Graduate School of Science and Engineering, Kinki University, 3-4-1 Kowakae, Higashi-Osaka, Osaka 577-8502, Japan}
\affiliation{Department of Physics, Azarbaijan University of Tarbiat Moallem, 53714-161, Tabriz, Iran}

\author{Mikio Nakahara}
\email{nakahara@math.kindai.ac.jp}
\affiliation{Research Center for Quantum Computing, Interdisciplinary Graduate School of Science and Engineering, Kinki University, 3-4-1 Kowakae, Higashi-Osaka, Osaka 577-8502, Japan}
\affiliation{Department of Physics, Kinki University, 3-4-1 Kowakae, Higashi-Osaka, Osaka 577-8502, Japan}

\begin{abstract}
We present a Lie--algebraic classification and detailed construction of the dynamical invariants, also known as Lewis--Riesenfeld invariants, of the four--level systems including two--qubit systems which are most relevant and sufficiently general for quantum control and computation. These invariants not only solve the time--dependent Schr\"odinger equation of four--level systems exactly but also enable the control, and hence quantum computation based on which, of four--level systems fast and beyond adiabatic regimes.
\end{abstract}

\maketitle

\section{Introduction}
Quantum computation is an emerging
discipline in which quantum physics is used as a computational
resource \cite{Nielsen2000,Nakahara2008}. By making use of states and operations,
which have no classical counterparts, a quantum computer is
expected to execute some hard tasks for a classical computer
in a reasonable time. Here operations are mostly elements of
SU($2^n$), where $n$ is the number of qubits, on which the
operations act.
It is known that any classical logic operation may be
realized by a collection of the NAND gate. The corresponding
``universality theorem'' is due to Barenco {\it et al.}
\cite{Barenco1995}. The theorem claims that ``any unitary gate
can be decomposed into one--qubit (i.e., SU(2)) gates and the CNOT
gates. In other words, the set of one--qubit gates and the CNOT
gate are universal in gate implementations. In many physical
systems, implementation of a one--qubit gate is often not hard.
It may be realized by the Rabi oscillation or the Raman
transition, for example. In contrast, implementation of the
CNOT gate can be challenging and its realization is sometimes regarded
as a milestone for a physical system to be a true candidate of
a working quantum computer \cite{DiVincenzo2000}.
Later, it turned out that any SU(4) gate, which entangles
a tensor product state, may serve as an element of a universal
set of quantum gates with the set of one--qubit gates
\cite{Barenco1995a,Lloyd1995}. Important exceptions of two--qubit
gates that are excluded are the SWAP gate and
the ``local gates'' SU(2)~$\otimes$~SU(2).

The above observations make the importance of implementation of an
SU(4) gate obvious. Adiabatic
two--qubit gate implementation is limited in time by the coupling
strength between the two qubits on which the gate acts.
In NMR, for example, the coupling strength $J$ is on the order
of $1 \sim 100$~Hz, leading to the execution time
on the order of millisecond to second, while the execution time
of a one--qubit gate is limited by the strength of the RF pulse
and it is typically on the order of $10~\mu$s for a
$\pi$--pulse for a heteronuclear molecule. It may take longer
for a homonuclear molecule. By considering this large difference
between the execution times, two--qubit gates often become a
bottleneck in shortening the execution time of a whole quantum circuit.
This is the motivation of considering non--adiabatic implementation of
nontrivial SU(4) gates.

Furthermore, there are many
nontrivial quantum algorithms, such as the Deutsch Algorithm,
the Grover algorithm and the Bernstein--Vazirani algorithm, just
to name a few, which can be demonstrated with a two--qubit system
and an SU(4) gate. Execution of these algorithms with a speed
beyond the adiabatic limit shows the promising future of
a realization of quantum computing.

\subsection{Dynamical Invariants}
An alternative way of constructing solutions of the time--dependent Schr\"odinger equation and obtaining the time--evolution operator is by means of the eigenstates of an operator $I=I^\dagger$ \footnote{We will use $I$ to denote a dynamical invariant and $\openone$ to denote $2 \times 2$ identity matrix throughout the paper.}, which is a dynamical invariant of the system with a time--independent expectation value $\braket{I}$ \cite{Lewis1969}. A dynamical invariant (DI) or a Lewis--Riesenfeld Invariant (LRI) obeys
\be
\langle I \rangle = \langle \psi(t)|I|\psi(t)\rangle = \langle\psi(0)| U^\dagger I U|\psi(0)\rangle = \mathrm{const.},
\ee
where $ U =  U(t;0)$ is the time evolution operator.
Using Heisenberg's equation (in units such that $\hbar=1$) we can restate this condition as
\be
0 = \frac{d}{dt}I^{(H)} = \left(\frac{\partial I}{\partial t}\right)^{(H)} + i[H^{(H)},I^{(H)}].
\label{eq:Heisenberg}
\ee
where the superscript $(H)$ denotes a Heisenberg picture operator $O^{(H)} =  U^\dagger O   U$, where $O$ is a Schr\"odinger picture operator. With the LHS being $0$, \eq{eq:Heisenberg} simplifies to
\be
0 = \frac{\partial I}{\partial t} + i[H, I],
\label{eq:Lewis1969}
\ee
which is an equation in the Schr\"odinger picture. We will refer to this Liouville--von Neumann type equation as the \emph{DI equation}. In terms of the eigenstates of $I$, $\ket{\phi_n(t)}$, the general solution of the time--dependent Schr\"odinger equation can be written as
\be
|\Psi(t)\rangle = \sum_n c_n e^{i \alpha_n(t)}|\phi_n(t)\rangle
\ee
and the time--evolution operator becomes
\be
 U(t;0) = \sum_n e^{i \alpha_n(t)}|\phi_n(t)\rangle \langle \phi_n(0) |
\label{eq:U}
\ee
where Lewis--Riesenfeld phase $\alpha_n(t)$ is given as \cite{Lewis1969}
\be
\alpha_n(t) = \int_0^t \bra{ \phi_n(s)} \left(i\frac{\partial}{\partial s} - H(s) \right) \ket{\phi_n(s)} ds. 
\ee
We observe that eigenstates of $I$ evolve in the following simple form
\be
e^{i \alpha_n(t)}|\phi_n(t)\rangle =  U(t;0)|\phi_n(0)\rangle.
\label{eq:phievolution}
\ee
This passage is transitionless in the eigenbasis of $I$, and is not necessarily adiabatic. The power of DIs goes beyond solving the time--dependent Schr\"odinger equation however. A major problem in adiabatic quantum control is that in many cases the evolution is so slow that the system may start decohering. Evolution in the eigenstates of $I$, however, is not restricted by the adiabaticity condition and can be made to be fast, a feature which caused a recent surge of applications \cite{FWN2011,Chen2010prl,ChenPRA2010,Chen2010a,ChenPRA2011a,Chen2011a,ChenPRA2011b,ChenPRA2011c,ChenPRA2011d}. DIs have been used in the context of quantum field theory \cite{Pedrosa2007}.
Another attractive feature of DI is that once obtained, the time--evolution operator of the system can be constructed from its eigenvectors following \eq{eq:U} (which can be otherwise obtained by the direct evaluation of the time--ordered integral \cite{Sakurai2010} or through Wei--Norman expansion \cite{Wei1963,Wei1964,Rau1998,Rau2000,Rau2005,Uskov2006,Uskov2008}).

\subsection{Quantum Control Based on DIs}
Quantum control is vital to quantum computation. For example, in one of the major quantum computation schemes, the AQC \cite{AQCsci2001,SarandyAQCOSprl2005}, a quantum gate is essentially an adiabatic quantum control passage. Nevertheless, in general, an adiabatic control of a system is very slow relative to the decoherence of the system  which may render an AQC model over the system impractical. To resolve this issue, a quantum computation scheme based on DIs has also been proposed \cite{Sarandy2011}. Nonetheless, two limitations hinder the quantum control methods and hence the quantum computation models based on DIs are not widely applicable, due to the three difficulties in obtaining a DI of a Hamiltonian.

In principle, a Hamiltonian of a system is a time--dependent operator on the Hilbert space of the system, so is a DI associated with the Hamiltonian; they satisfy the DI equation. The first two difficulties regard the operator form of the DIs of a Hamiltonian. A Hamiltonian can spawn multiple DIs, as aforementioned; hence, one must not only know all possible DIs of a Hamiltonian but also select from among them the one pertinent to the end of the control, and determine whether the choice is optimal. These difficulties worsen dramatically with the system's level. Unfortunately, the current literature has not been able to cope with these difficulties; instead, it usually fumbles a way to certain DIs. Note that in the case of two--level, however, a classification of Hamiltonians and their DIs does exist \cite{ShenLRI2Exact2003}. Consequently, insofar the quantum control and computation models  based on DIs are nigh limited to two--level systems \cite{FWN2011,Chen2010prl,ChenPRA2010,Chen2010a,ChenPRA2011a,
Chen2011a,ChenPRA2011b,ChenPRA2011c,ChenPRA2011d}. Special cases in four--level systems have also been studied \cite{Sarandy2011}. 

Even if the operator form of a DI is known (it may or may not belong to a subalgebra of $\su(4)$ as we detail in Section \ref{sec:obtaining}) it is still hard to obtain the DI in closed--form by solving the DI equation, which is in fact a set of differential equations, the number of which depends on the number of degrees of freedom of the system. Besides partially causing the first limitation, this third difficulty circumscribes how and how much one can control a quantum system based on the DIs of the system. In fact, because of this limitation, in order to gain the control of a quantum system via a DI of the system, a special method ---the inversely engineered control (IEC)--- is needed, which was invented in \cite{Chen2011a} for pure states and extended to mixed states in \cite{FWN2011}.   

Although this paper is not about quantum control, so as to understand some settings for classifying DIs in the main text, let us briefly introduce the gist of quantum control, in regard of AQC and IEC. To control a quantum system one needs to expose the system in an external electromagnetic field that interacts with the system, such that the total Hamiltonian of the system and the external field evolves temporally in the way that takes the system to a designated state at some point of time. A system under an AQC remains in an instantaneous eigenstate of the total Hamiltonian that must satisfy the adiabaticity condition, the violation of which would render the AQC of the system impossible.

In the control based on a DI of a system, however, the system does not necessarily follow any instantaneous eigenstate of the total Hamiltonian of the system and the external field but rather always stays in an instantaneous eigenstate of the DI. This idea of control is realized by the IEC method. In contrast to the logic that one should know a Hamiltonian first to determine its DIs, in IEC, one only get hold of the matrix form of a Hamiltonian, while leaving its physical parameters unknown, assume the matrix form of a DI and postulate some simple functional forms of its parameters, determine the physical parameters of the DI by the boundary conditions that comply with the wanted initial and final states of the control, and then use the DI equation to nail down the Hamiltonian that bears this DI.
In doing so, one avoids the difficulty of solving the DI equation directly from a known Hamiltonian. Aside of this mathematical advantage of IEC, a physical advantage is that the 
control can be done regardless of the adiabaticity condition and thus serves as a shortcut to adiabatic control. But the costs are: 1) the Hamiltonian obtained this way may be unphysical or impractical 2) a great deal of control range is lost due to the restricted postulates one can make of the parameters of the DI.

In \cite{Chen2011a} the method is demonstrated in an $\su(2)$ problem. The DI is taken to be a function of the form
\be
I=\frac{\Omega_0}{2} \left(\sin\gamma\cos\beta \sigma_x + \sin\gamma\sin\beta \sigma_y +\cos\gamma\sigma_z\right),
\ee
where $\gamma$ and $\beta$ are polynomials in $t$ of order 3 and 4 respectively, by ansatz. The system is assumed to be prepared in a shared eigenstate of $I$ and $H$ initially, and due to \eq{eq:phievolution} there will not be any transitions in the eigenbasis of $I$. This situation is similar to adiabatic quantum control where the passage is transitionless in Hamiltonian's eigenbasis, but without the adiabaticity condition.
The corresponding Hamiltonian is found by using the DI equation as
\be
H=\frac{1}{2}\left[ \frac{\dot \gamma}{\sin \beta} \sigma_x + \left( \frac{\dot \gamma}{\sin \beta} \cot\gamma \cos\beta -\dot\beta \right) \sigma_z \right].
\label{eq:HChen2011a}
\ee
The coefficients of the polynomials are fixed by employing the population inversion condition at a final time $t_f$, and requiring $[H(0),I(0)]=[H(t_f),I(t_f)]=0$ which ensures that the initial and final states are eigenstates of the Hamiltonian as well. $t_f$ can be arbitrarily small, as long as it does not violate the time--energy uncertainty relation \cite{Chen2011a,Sarandy2011}, resulting in a passage that can be made fast. Unfortunately, a precise implementation of the required Hamiltonian is not an easy task.

One may wonder if in IEC it is still difficult to guess the matrix form of a DI of a Hamiltonian. Indeed, it is still difficult; however, as IEC has been applied to only two--level systems so far \footnote{In \cite{Sarandy2011}, where a quantum computation model based on DIs are proposed, the control method of four--level systems is essentially the same as IEC.}, the difficulty is negligible, as one can always cast the DI in a linear combination of the three Pauli matrices, which is simple enough and without loss of generality. A four--level system has $15$ generators, rendering the operator form of DI a lot more difficult to guess and resulting set of differential equations are less likely to be solved analytically. A companion paper addresses the issue by offering a Lie algebraic classification of DIs for Hermitian, finite--level systems \cite{Wan2012}. To this end, we assume in this paper four--level Hamiltonians and their DIs live in the Lie algebra $\su(4)$ as discussed in \cite{Wan2012}. Below we list 
our 
main results.
\begin{enumerate}[label=\roman*)]
 \item We apply the key ideas addressed in \cite{Wan2012} to four--level systems which are of particular importance in quantum computation.
 \item We show DIs of four--level systems can be classified in terms of the maximal subalgebras of $\su (4)$.
 \item By means of this classification, we construct DIs for a family of four--level system, which are sufficiently general for applications.
 \item  Our classification indicates a great reduction of the complexity in constructing exact DIs of four--level systems.
 \item We list explicit set of differential equations for both types, and further discuss the cases which are both exactly solvable and physically feasible.
\end{enumerate}

This paper is organized as follows.
In the next section, we give a review of the two types of DIs and the adjoint DI equation. We move on to the classification of $\su(4)$ DIs with respect to the subalgebra that spans the set of generators in the Hamiltonian in the following section, and give detailed analysis for each possible case. Finally we summarize our results. Throughout the rest of the paper, summation over repeated indices is assumed unless stated otherwise.

\section{Obtaining Dynamical Invariants}
\label{sec:obtaining}
We summarize the relevant findings of \cite{Wan2012} in a self--contained manner from an $\su(4)$ point of view in this section. A given Hamiltonian can have infinite number of DIs. Since $I \in \su(4)$ we can expand it linearly in terms of a convenient set of $\su(4)$ generators $\mathfrak I$ with time--dependent coefficients as $g_i^{(\lambda)} \lambda_i = g_i^{(\lambda)}(t) \lambda_i$. (For brevity, we will drop the superscript denoting the representation for the fundamental representation and set $g_i = g_i^{(\lambda)}$ in what follows). We assume $H \in \su(4)$, since we can always drop the identity element which would otherwise commute with everything per \eq{eq:Lewis1969} and result in the same equation. Letting $\mathfrak H$ be the set that generates $H$ and ignoring the coefficients $i f_{ijk}$ for a moment (we use the convention in which the structure constants obey $[\lambda_i, \lambda_j] = i f_{ijk} \lambda_k$), due to \eq{eq:Lewis1969} $\mathfrak I$ will obey
\be
\mathfrak I = [\mathfrak H, \mathfrak I].
\label{eq:Lewis1969Set}
\ee
When we choose to include any generator from the set $\mathfrak H$ in $\mathfrak I$, as a consequence of Cartan decomposition, $\mathfrak I$ spans the minimal subalgebra $\mathfrak S$ that encloses $\mathfrak H$, with $\mathfrak H \subseteq \mathfrak I$, $\mathrm{span}(\mathfrak I) = \mathfrak S \subseteq \su(n)$. We refer this class as \emph{S(uperset)--type DI}, or for the purposes of this paper equivalently \emph{S(ubalgebra)--type DI}, which has been the focus in the literature so far.

Another class of DIs are generated by a completely disjoint set of generators from that of $\mathfrak H$. As we will see in Section \ref{sec:fourlevel}, the number of DIs obeying $\mathfrak H \cap \mathfrak I = \emptyset$, which we will refer as \emph{D(isjoint)-type DIs}, is determined by the embeddings of the subalgebras of $\su(n)$, once the minimal enclosing subalgebra of $\mathfrak H$ is determined. Clearly when $\mathfrak S = \su(n)$, there are no D--type DIs. We will see that in some cases, such as $\mathfrak S = \so(5) \subset \su(4)$, a D--type invariant can simplify the problem.

When we plug $I=g_i \lambda_i$ into the DI equation, we obtain
\be
 0 = \dot{\vect g} + i A \vect g,
\label{eq:adjoint}
\ee
hereafter called the \emph{adjoint DI equation}, where $\vect g = (g_{1}, \ldots, g_{15})^T$ is a vector made of the expansion coefficients of $I$. At first sight, $i A$ is a $15 \times 15$ anti--symmetric, real matrix whose entries are linear combinations of the coefficients in $H$. 
By casting \eq{eq:Lewis1969} into
\be
0 = \dot g_i \lambda_i + g_j [H, \lambda_j ]
\ee
and comparing with \eq{eq:adjoint}, we see that the $A$ matrix can be obtained from the Hamiltonian directly as $\ad(H)$. This means $A$ matrix lives in the adjoint representation of $\su(4)$, that is $A \in \ad [\su(4)]$ in general. $A$ obtained this way will be a block diagonalized $15 \times 15$ matrix, with different blocks representing S--type and D--types (if there is any). Once the adjoint representation is known, this serves as a very practical way of obtaining $A$.

Due to the DI equation, if $I$ is a valid DI and if a Hermitian time--independent operator $\Lambda$ is such that $[H, \Lambda]=0$, then $I' = c_1 I + c_2 \openone + c_3 \Lambda$ where $\{c_i\}$ are constants is also a valid DI. This freedom allows one to work with a ``minimal'' DI in which the time--independent trace part and terms that commute with $\mathfrak H$ are dropped. We will exploit this freedom in what follows, and work with this kind of minimal DIs. We remark that since $I$ obeys the same Liouville--von Neumann equation just like the density matrix $\rho$, it is possible to obtain a density matrix from a DI vice and versa, some examples given in \cite{Ichiyanagi1986,Saradzhev2007}. Traceless part of $\rho$ is always a valid DI, but the converse is not true. A density matrix has trace one and is positive semi--definitive whereas a DI is not necessarily, however this situation may be remedied using a ``non--minimal'' DI when necessary.

\section{Four--Level Systems}
\label{sec:fourlevel}
In this paper, we will be working with a form of four--level Hamiltonian which is general enough to cover almost all practical applications \footnote{By doing that we leave out some systems such as ENDOR involving anisotropy with electron spin and nuclear spin coupling \cite{Sato2007}. }
\be
J_i \sigma_i \otimes \sigma_i + h^{(1)}_i \sigma_i \otimes \openone + h^{(2)}_i \openone \otimes \sigma_i \quad (i=x,y,z)
\label{eq:H}
\ee
with at least one non--zero $J_i$. Each $J_i$ is a coupling coefficient, and each $h^{(1)}_i$ ($h^{(2)}_i$), is a field acting on the first (second) qubit, and is sometimes referred to as a control parameter. There could be an extra term $\openone \otimes \openone$ in the Hamiltonian which we deliberately dropped; it would commute with everything else and would not appear in the adjoint DI equation. We will choose to drop any generator that commutes with all the others for the same reason. For purposes of quantum control we prefer to work with a respresentation involving elements of the form $\{\sigma_i \otimes \openone,\openone \otimes  \sigma_i,\sigma_i \otimes \sigma_j\}$ which allows us to read off the physical content directly. Moreover, this representation is compatible with the spinor representation of $\so(4)$ and $\so(5)$, and will be referred as such from here on. Another advantage of this spinor basis is that the commutator of two spinor elements is never a linear combination of more than one 
generators, which 
lets us use \eq{eq:Lewis1969Set} smoothly.
We will use the shorthand notation $XY$ to denote $\sigma_x \otimes \sigma_y$ interchangeably throughout the paper.

 The $A$ matrix corresponding to the subalgebra (S--type) and disjoint (D--type, may or may not exists) cases can be read off directly as separate blocks after block--diagonalization, in which case $\vect g$ belongs to one of the two subspaces that correspond to two distinct DIs. There can be more than one S--type or D--type DIs which will cause further block diagonalization in the respective sector. The overall block structure and the algebraic structure are determined through the embedding of the corresponding subalgebra of the adjoint representation which is the $15$ dimensional representation of $\su(4)$ in our case. Possible group embeddings are listed in Table \ref{tab:embeddings}.

\begin{table}
 \begin{ruledtabular}
\begin{tabular}{l | c}
$\su(3) \oplus \mathfrak u(1)$ & $3_{-4} + \bar 3_4 + 1_0 + 8_0$ \\
$\so(4)$ & $(3,1) + (1,3) + (3,3)$ \\
$\so(4) \oplus \mathfrak u(1)$ & $(1,1)_0 + (3,1)_0 + (1,3)_0 + (2,2)_2 + (2,2)_{-2}$ \\
$\so(5)$ & $5+ 10$ \\
\end{tabular}
 \end{ruledtabular}
 \caption{Maximal subalgebras of $\su(4)$ in the adjoint representation \cite{Slansky1981}.}
\label{tab:embeddings}
\end{table}

In the spinor basis, $\mathrm{span}(\mathfrak I)$ can be $\so(5)$, $\so(4) \oplus \mathfrak u (1)$ and $\so(4)$ subalgebras of $\su(4)$. It can not, however, be $\su(3) \oplus \mathfrak{u}(1)$ in this representation; one can constraint the coefficients of the generators to effectively transform the representation into a suitable one but the constraints are found to be too tight when we force the ``cross--terms'' such as $XY$ and $ZX$ to be zero in order to adapt \eq{eq:H}: the number of generators reduces to four, leaving us in $\su(2) \oplus \mathfrak u (1) \subset \su(3) \oplus \mathfrak u (1)$ (see Appendix \ref{sec:nosu3} for details). $\so(4) \oplus \mathfrak u (1)$ subalgebra arises from Hamiltonians that can have ---but not necessarily present--- a $\mathfrak u (1)$ generator representing a non--interacting subsystem, however because we choose ignore such triviality from the beginning we will not have the $\mathfrak u (1)$ generator in the invariant. 

While we can obtain $A$ directly by taking the adjoint of $H$ which is practical, we may desire to switch to a different choice of generator basis afterwards, such as the spinor basis \footnote{Because there is no standard ordering for the spinor basis and we will need to use different orderings later on, we do not list the structure constants for them.} with a particular ordering or a new set of their linear combinations. In this case, we may simply perform a time--independent similarity transformation on the adjoint DI equation: let $\{\Sigma_i\}$ be the new basis for DI such that $I= g_i \lambda_i = g^{(\Sigma)}_i \Sigma_i $. The transformation matrix relating two sets of generators through $\Sigma_i = [S]_{ij}\lambda_j$ can be obtained using $\tr(\lambda_i \lambda_j)=2\delta_{ij}$ as $[S]_{
ij}=\tr(\Sigma_i \
\lambda_j)/\tr(\lambda_i \lambda_i)$. Note that this mixes the original generators rather than transforming them altogether in the same way. The $\vect g$ vector will transform the same way as $\vect \lambda$ does, since $g_i \lambda_i$ as a whole forms as a single ``component'' of $I$, 
from which we deduce that 
the adjoint DI equation will become
\be
\begin{aligned}
 0 &= \dot {\vect g}^{(\Sigma)} + i A^{(\Sigma)} \vect g^{(\Sigma)}; \\ 
\qquad A^{(\Sigma)} &= S A S^{-1}, \quad \vect g^{(\Sigma)} = S \vect g.
\end{aligned}
\label{eq:DItransform}
\ee
Clearly, for the differential equation
\be
 0 = \frac{\partial}{\partial t} (S \vect{g}) + i(S A S^{-1}) (S \vect{g})
\ee
to keep its form after the similarity transformation, we must have $(\partial S/\partial t) \vect{g} = 0$, which is always the case when the new frame is stationary as in our case.

We observe that the choice of axes is not of any importance, as any similarity transformation on the set of generators as a whole leaves the subalgebra spanned by $\mathfrak I$ and the structure constants intact, resulting in the same equation as above. Nevertheless, the choice of generator basis does make a difference by mixing the components of $\vect g$ and determining the block structure of $A$.

\subsection{Classification of Hamiltonians and Invariants}
Let us denote an S--type $\mathfrak I$ with $\mathfrak I_S$ and a D--type $\mathfrak I$ with $\mathfrak I_D$. Given a Hamiltonian, the subalgebra spanned by $\mathfrak I_S$ can be determined directly by means of Table \ref{tab:subalgebras}. When $|\mathfrak H| \leq 3$, due to the simplicity, one can manually check whether these generators belong to an $\su(2)$. In other cases, finding the smallest set in the table such that $\mathfrak H$ fits into gives the answer. However, we can alternatively decide on the desired subalgebra as our starting point and pick a subset of corresponding generators from the table to form a Hamiltonian, which will be our approach in this paper to exhaust all possibilities in a concise way.

Depending on the list of generators appearing in $\mathfrak I_S$, further embeddings are possible through the embedding of the subgroups listed in Table \ref{tab:subalgebras}. Note that $\so(4) \subset \su(4)$ and $\so(4) \subset \so(4) \oplus \mathfrak u(1) \subset \su(4)$ will result in different D--types and different block structures for $A$. The block structure is dictated by the maximal embedding.

\begin{table}
\begin{ruledtabular}
\begin{tabular}{l | c }
$\su(3) \oplus \mathfrak u(1)$ & $-$ \\
\hline
$\so(4)$ & $\{ \sigma_i \sigma_j, \sigma_j \sigma_j, \sigma_k \openone, \sigma_k \sigma_i, \sigma_k \sigma_k, \openone  \sigma_j \}$. \\
\hline
$\so(4) \oplus \mathfrak u (1)$ & $\{\openone \sigma_*, \sigma_i \sigma_* \} \sqcup \{ \sigma_i \openone  \}$, \\
& $\{   \sigma_i  \openone,   \openone \sigma_j,   \sigma_j \sigma_k,  \sigma_j \sigma_i, \sigma_k \sigma_i, \sigma_k \sigma_k \} \sqcup \{ \sigma_i \sigma_j  \}$, \\
& $\{ \sigma_i \openone , \openone  \sigma_i, \sigma_j \sigma_k, \sigma_j \sigma_j, \sigma_k \sigma_j, \sigma_k \sigma_k \} \sqcup \{\sigma_i \sigma_i \}$.  \\
\hline
$\so(5)$ & $\{  \sigma_i \openone, \openone \sigma_*, \sigma_j \sigma_*,  \sigma_k \sigma_*  \}$. \\
\end{tabular}
\end{ruledtabular}
\caption{List of maximal (semi--)simple subalgebras $\mathfrak S \subset \su(4)$ when $\mathfrak H \subseteq \{\sigma_* \openone,  \openone \sigma_*, \sigma_x \sigma_x, \sigma_y\sigma_y, \sigma_z \sigma_z\}$ with at least one $\sigma_i \sigma_i$ type generator, in accordance with \eq{eq:H}. There is no summation over repeated indices. The indices $i,j,k \in \{x,y,z\}$ are distinct. $\sigma_*$ serves as a wildcard in such a way that $\sigma_i \sigma_*$ denotes three elements $\sigma_i \sigma_x,\sigma_i \sigma_y,\sigma_i \sigma_z$. The complementary table is obtained by swapping the role of first and second terms in all generators. Alternative tables can be obtained by similarity transformations. Tensor product symbol is omitted for brevity. The table is referred and used in Section \ref{sec:so4pu1}.
}
\label{tab:subalgebras}
\end{table}

\subsection{Exact Dynamical Invariants}
The adjoint DI equation for the $\su(2) \cong \so(3)$ case has been studied in the literature extensively under different contexts, with many known exactly solvable cases, and is known as the Bloch equation with infinite relaxation \cite{Rosen1932,Silver1985,Grunbaum1986,Town1989,Prants1990,Prants1990a,Shinnar1993,Rosenfeld1996,Bagrov2001,Barata2002,Kobayashi2002,Kobayashi2004a,Kobayashi2004b,Bagrov2005,Brouder2007,Rourke2007}. Thus the Bloch equation is a special case of the adjoint DI equation.

In both $\so(4) \cong \su(2) \oplus \su(2)$ and $\so(4) \oplus \mathfrak u(1)$ cases we essentially have two decoupled Bloch equations, meaning that the mathematical problem reduces to solving single--qubit problems. Note that the two physical qubits are actually coupled in $\so(4) \oplus \mathfrak u(1)$ case but not in $\so(4)$ case.  The explicit block diagonalization to bring $(3,1) + (1,3)$ into $3+3$ can be achieved by a time--independent similarity transformation. To this purpose, we first notice that $\so(4)$ is readily split into two distinct $\su(2)$, $\{ \openone  \sigma_* \} \sqcup \{ \sigma_* \openone\}$ (note that some generators may require a correction of sign depending $\{i,j,k\}$ being an odd permutation of $\{x,y,z\}$ or not). Clearly, such a Hamiltonian represents composite 
systems with two non--interacting $\su(2)$ subsystems. The other $\so(4)$ case  $\{ \openone \sigma_j, \sigma_k \sigma_i, \sigma_k \sigma_k \} \sqcup \{ \sigma_k \openone, \sigma_i \sigma_j, \sigma_j \sigma_j \}$ is equivalent to the former case up to a similarity transformation.
Moreover, the D--type DI for $\so(4)$ results in a $9\times 9$ $A$ matrix, which does not help simplifying the problem. It is because of these facts that we will not analyze $\so(4)$ subalgebra any further \footnote{Although the physical context is different, \cite{Suchowski2011} studies an $\so(4)$ Hamiltonian in a Lie--algebraic framework.}. In $\so(4) \oplus \mathfrak u (1)$ cases however, we respectively 
take the following simple linear combinations of the generators to form two 
separate $\su(2)$ algebras
\be
\begin{aligned}
\{(\openone \pm \sigma_i )\sigma_*\}/2, \\
\{ \sigma_j \openone \pm \openone \sigma_k, \sigma_i \sigma_j \mp \sigma_k \sigma_i,  \sigma_k \sigma_j \pm \sigma_i \sigma_i\}/2,  \\
\{ \sigma_i \openone \pm \openone \sigma_i, \sigma_j \sigma_j \mp \sigma_k \sigma_k,  \sigma_j \sigma_k \pm \sigma_k \sigma_j\}/2,
\label{eq:su2psu4}
\end{aligned}
\ee
in the notation and order of Table \ref{tab:subalgebras}. Note that the summands in each new generator are related to each other  by the $\mathfrak u (1)$ generator of the subalgebra; we have exploited the following to obtain these pairs: let $\{T_1,T_2,T_3\} \sqcup {q}$ (where $\{T_1,T_2,T_3\} \subset \mathfrak I_S$ and $q^2=1$) be an $\su(2) \oplus \mathfrak u(1)$, then $\{T_1 \pm q T_1,T_2 \pm q T_2,T_3 \pm q T_3\}/2$ is a pair of $\su(2)$ and $\{T_1,T_2,T_3, q T_1,q T_2,q T_3\} \sqcup \{q\}$ is an $\so(4) \oplus \mathfrak u(1)$.
Different choices such as the one given in \cite{Rau2000,Rau2005} can be obtained through time--independent similarity transformations.
Furthermore, the sector corresponding to the D--type DI further splits into two $4 \times 4$ blocks in $\so(4) \oplus \mathfrak u(1)$, resulting in four smaller, decoupled adjoint DI equations. These two sets of $4$ generators are conjugate of each other under the respective $\mathfrak u (1)$ generator.
There are three types of $\so(4) \oplus \mathfrak u(1)$ (see Table \ref{tab:subalgebras}) which we will discuss in detail below.

\subsubsection{$\so(4) \oplus \mathfrak u (1)$}
\label{sec:so4pu1}
\textbf{1.} This first case is also known as single--qubit control Ising model, and the corresponding Hamiltonian can be written as
\be
H = J XX + h_x \openone X + h_y \openone Y + h_z \openone Z.
\label{eq:HIsing}
\ee
A quick comparison of the set of generators $\mathfrak H = \{\openone X, \openone Y, \openone Z, XX \}$ of this Hamiltonian with Table \ref{tab:subalgebras} reveals $\mathrm{span}(\mathfrak H) \subset \so(4) \oplus \mathfrak u (1)$ (the first entry in the table, due to the distinctive control terms $\openone \sigma_*$, along with $\sigma_i \sigma_i$), with the algebraic groupings given as
\be
\label{eq:chargeconjugation}
\begin{aligned}
\{\Sigma_i\} = \{\openone X, \openone Y, \openone Z, XX,XY,XZ\} \sqcup \{X \openone \} \sqcup  \\
  \{ZX, ZY, ZZ, Y \openone\} \sqcup \{YX,YY,YZ,-Z\openone\} \\
= \mathfrak I_S \sqcup Q \sqcup \mathfrak I_D \sqcup \mathfrak  I_{\bar D}.
\end{aligned}
\ee
The sets $\mathfrak I_D$ and $\mathfrak I_{\bar D}$ representing two D--type invariants obey \eq{eq:Lewis1969Set}, and are related to each other by an infinitesimal $\mathfrak u (1)$ charge conjugation
\be
\[Q, \mathfrak I_D\] =  -2 i \mathfrak I_{\bar D}, \qquad \[Q, \mathfrak I_{\bar D}\] = 2 i \mathfrak I_{D}.
\ee
Exploiting the $\mathfrak u (1)$ factor of the subalgebra, it is possible to consider an additional time--dependent $XI$ term in the Hamiltonian as well, but it will not affect the discussion that follows.
Although the $(3,1)+(1,3)$ $\so(4)$ sector is not two readily decoupled $\su(2)$ systems, such a situation can always be remedied by means of a time--independent similarity transformation which we will employ below. This is of course not the case for the $(2,2)_{\pm 2}$ sector.

In the generator basis $\{\Sigma_i\}$ the adjoint DI equation is $0 = \vect {\dot g}^{(\Sigma)} + i A^{(\Sigma)} \vect g^{(\Sigma)}$ and $A^{(\Sigma)} $ has the block diagonal form $A^{(\Sigma)} = A^{(\Sigma)}_{(3,1)+(1,3)} \oplus (0) \oplus A^{(\Sigma)}_{(2,2)_{-2}} \oplus A^{(\Sigma)}_{(2,2)_{2}}$ where
\be
A^{(\Sigma)}_{(3,1)+(1,3)}  &=& 
2i \left(
\begin{array}{cccccc}
 0 & -h_z & h_y & 0 & 0 & 0 \\
 h_z & 0 & -h_x & 0 & 0 & -J \\
 -h_y & h_x & 0 & 0 & J & 0 \\
 0 & 0 & 0 & 0 & -h_z & h_y \\
 0 & 0 & -J & h_z & 0 & -h_x \\
 0 & J & 0 & -h_y & h_x & 0 \\
\end{array}
\right), \nonumber　\\
A^{(\Sigma)}_{(2,2)_{\pm 2}} &=& 2i\left( \begin{array}{cccc}
 0 & -h_z & h_y & J \\
 h_z & 0 & -h_x & 0 \\
 -h_y & h_x & 0 & 0 \\
 -J & 0 & 0 & 0
\end{array} \right).
\ee
and $(0)$ is a $1\times 1$ matrix containing $0$. The corresponding linear combinations of $\so(4)$ generators listed in \eq{eq:su2psu4} are the two independent sets of $\su(2)$, $\{ \Sigma_i^\pm\} = \{(\openone \pm X)X, (\openone \pm X)Y, (\openone \pm X)Z \}/2$.
Transforming to the basis $\{\Sigma'_i\} = \{ \Sigma^+\} \sqcup \{ \Sigma^- \}$ by employing \eq{eq:DItransform} with $[S]_{ij} = \tr(\Sigma'_i \Sigma_j)/4$,
\be
S=\frac{1}{2}\left(
\begin{array}{cccccc}
 1 & 0 & 0 & 1 & 0 & 0 \\
 0 & 1 & 0 & 0 & 1 & 0 \\
 0 & 0 & 1 & 0 & 0 & 1 \\
 1 & 0 & 0 & -1 & 0 & 0 \\
 0 & 1 & 0 & 0 & -1 & 0 \\
 0 & 0 & 1 & 0 & 0 & -1
\end{array}
\right)
\ee
will block diagonalize $A^{(\Sigma)}_{(3,1)+(1,3)}$, yielding
\be
0 = \dot {\vect g}^\pm + i A^\pm \vect g^\pm 
\ee
where
\be
\begin{aligned}
A^\pm &= 
-2i\left(
\begin{array}{ccc}
 0 & -h_z & h_y \\
 h_z & 0 & -(h_x \pm J) \\
 -h_y & h_x \pm J & 0
\label{eq:so4pu1Apm1}
\end{array}
\right),  \\	
\vect g^\pm &= (g_{\openone X} \pm g_{XX},g_{\openone Y} \pm g_{XY}, g_{\openone Z} \pm g_{XZ})^T/2.
\end{aligned}
\ee

\textbf{2.} The second type of Hamiltonian in $\so(4) \oplus \mathfrak u (1)$ can be written as
\be
H = J XX + h^{(1)} Y\openone  + h^{(2)} \openone Z.
\ee
Generators of the S--type, its respective $\mathfrak u(1)$ generator as well as the generators of the D--types are grouped as
\be
\begin{aligned}
 \{ \Sigma_i \} = \{ XX, Y\openone , ZX, ZY, \openone Z, XY \} \sqcup \{ YZ \} \sqcup \\
\{ X\openone , Z\openone , YY, YX \} \sqcup \{ ZZ, -XZ, \openone X, -\openone Y \}.
\end{aligned}
\ee
Taking the adjoint $A^{(\Sigma)}=\ad(H)$, we obtain $A^{(\Sigma)}_{(3,1)+(1,3)} \oplus (0) \oplus A^{(\Sigma)}_{(2,2)_{-2}} \oplus A^{(\Sigma)}_{(2,2)_{2}}$,
\be
A^{(\Sigma)}_{(3,1)+(1,3)} &=& 2i \left( \begin{array}{cccccc}
0  & 0 & h_y^{(2)} & 0 & 0 & -h_z^{(1)} \\
 0 & 0 & -J_x & 0 & 0 & 0 \\
 -h_y^{(2)} & J_x & 0 & -h_z^{(1)} & 0 & 0 \\
 0 & 0 & h_z^{(1)} & 0 & 0 & -h_y^{(2)} \\
 0 & 0 & 0 & 0 & 0 & J_x \\
 h_z^{(1)} & 0 & 0 & h_y^{(2)} & -J_x & 0
\end{array}\right) \nonumber \\
A^{(\Sigma)}_{(2,2)_{\pm2}} &=& 2i \left(\begin{array}{cccc}
 0 & h_y^{(2)} & 0 & 0 \\
 -h_y^{(2)} & 0 & 0 & J_x \\
 0 & 0 & 0 & h_z^{(1)} \\
 0 & -J_x & -h_z^{(1)} & 0
\end{array}\right)
\ee
Similar to the previous case, we note that $\{ XX \pm ZY, Y\openone  \pm \openone Z,  ZX \mp XY \}/2$ generates an $\su(2)$ algebra which enables us to further block diagonalize $A_{(3,1)+(1,3)}$ into
\be
A^\pm = 2i\left(\begin{array}{ccc}
 0 & 0 & h_y^{(2)} \pm h_z^{(1)} \\
 0 & 0 & -J_x \\
 -(h_y^{(2)} \pm h_z^{(1)}) & J_x & 0
\end{array}\right),
\label{eq:so5Apm1}
\ee
with
\be
\vect g^\pm = (g_{XX} \pm g_{ZY}, g_{Y\openone } \pm g_{\openone Z}, g_{ZX} \mp g_{XY})^T/2.
\ee

\textbf{3.} Finally, we consider the Hamiltonian
\be
H = J_x XX + J_y YY + h^{(1)} Z\openone  + h^{(2)} \openone Z.
\ee
which is the last instance of the $\so(4) \oplus \mathfrak u(1)$ listed in Table \ref{tab:subalgebras}.
This Hamiltonian can implement a two--qubit gate because it can represent Josephson junctions when one of the coupling terms $J_i$ vanishes \cite{Niskanen2005}; also with the addition of the $\mathfrak u(1)$ generator $ZZ$ and the assumption  $J_x=J_y=J_z$, the Hamiltonian can implement $D_3$ model \cite{Zhou2011,Zhou2012} and quantum dots \cite{Loss1998}.

The generators are grouped as
\be
\begin{aligned}
 \{ \Sigma_i \} = \{ XX, YX,Z\openone , YY, XY, \openone Z\} \sqcup \{ZZ\} \sqcup  \\ 
\{YZ, \openone Y, XZ, \openone X \}  \sqcup \{X\openone ,ZX, -Y\openone ,-ZY\},
\end{aligned}
\ee
 and $A^{(\Sigma)}$ is $A^{(\Sigma)}_{(3,1)+(1,3)} \oplus (0) \oplus A^{(\Sigma)}_{(2,2)_{-2}} \oplus A^{(\Sigma)}_{(2,2)_{2}} $ where
\be
A^{(\Sigma)}_{(3,1)+(1,3)} &=&
2i\left(
\begin{array}{cccccc}
 0 & -h^{(1)} & 0 & 0 & -h^{(2)} & 0 \\
 h^{(1)} & 0 & -J_x & -h^{(2)} & 0 & J_y \\
 0 & J_x & 0 & 0 & -J_y & 0 \\
 0 & h^{(2)} & 0 & 0 & h^{(1)} & 0 \\
 h^{(2)} & 0 & J_y & -h^{(1)} & 0 & -J_x \\
 0 & -J_y & 0 & 0 & J_x & 0
\end{array}
\right) \nonumber \\
A^{(\Sigma)}_{(2,2)_{\pm2}} &=& 2i\left( \begin{array}{cccc}
 0 & 0 & h^{(1)} & -J_y \\
 0 & 0 & -J_x & h^{(2)} \\
 -h^{(1)} & J_x & 0 & 0 \\
 J_y & -h^{(2)} & 0 & 0
\end{array} \right).
\ee
$\{ XX \mp YY, YX \pm XY, Z\openone  \pm \openone Z \}/2$ is an $\su(2)$ and $A_{(3,1)+(1,3)}$ decomposes into
\be
A^\pm &=& 2i\left( \begin{array}{ccc}
 0 & -(h^{(2)} \mp h^{(1)}) & 0 \\
  h^{(2)} \mp h^{(1)} & 0 & J_y \pm J_x \\
 0 &  -(J_y \pm J_x) & 0
\end{array}\right) \nonumber \\
\vect g^\pm &=& (g_{XX} \mp g_{YY}, g_{YX} \pm g_{XY}, g_{Z\openone } \pm g_{\openone Z})/2.
\label{eq:so5Apm2}
\ee

\subsubsection{$\so(5)$}
We now turn to the most general $\so(5)$ case of the Hamiltonian \eq{eq:H}, which also is the last case in Table \ref{tab:subalgebras}:
\be
\begin{aligned}
H =& J_x XX + J_y YY +  \\ 
& h_x^{(2)} \openone X + h_y^{(2)} \openone Y + h_z^{(2)} \openone Z + h_z^{(1)} Z\openone
\end{aligned}
\ee
This time, the generators are grouped into two sets per Table \ref{tab:embeddings}:
\be
\{\Sigma\} &=& \{\openone X, \openone Y, \openone Z, XX, XY, XZ, YX, YY, YZ, Z\openone \} \sqcup \nonumber \\
&& \{ZX, ZY, ZZ, X\openone , Y\openone \}
\ee
$A^{(\Sigma)}$ matrix is a direct sum of two blocks $A^{(\Sigma)}_{10} \oplus A^{(\Sigma)}_5$, where

\begin{widetext}
\be
\begin{aligned}
A^{(\Sigma)}_{10} &=
i \left(\begin{array}{cccccccccc}
 0 & -h_z^{(2)} & h_y^{(2)} & 0 & 0 & 0 & 0 & 0 & J_y & 0 \\
 h_z^{(2)} & 0 & -h_x^{(2)} & 0 & 0 & -J_x & 0 & 0 & 0 & 0 \\
 -h_y^{(2)} & h_x^{(2)} & 0 & 0 & J_x & 0 & -J_y & 0 & 0 & 0 \\
 0 & 0 & 0 & 0 & -h_z^{(2)} & h_y^{(2)} & -h_z^{(1)} & 0 & 0 & 0 \\
 0 & 0 & -J_x & h_z^{(2)} & 0 & -h_x^{(2)} & 0 & -h_z^{(1)} & 0 & J_y \\
 0 & J_x & 0 & -h_y^{(2)} & h_x^{(2)} & 0 & 0 & 0 & -h_z^{(1)} & 0 \\
 0 & 0 & J_y & h_z^{(1)} & 0 & 0 & 0 & -h_z^{(2)} & h_y^{(2)} & -J_x \\
 0 & 0 & 0 & 0 & h_z^{(1)} & 0 & h_z^{(2)} & 0 & -h_x^{(2)} & 0 \\
 -J_y & 0 & 0 & 0 & 0 & h_z^{(1)} & -h_y^{(2)} & h_x^{(2)} & 0 & 0 \\
 0 & 0 & 0 & 0 & -J_y & 0 & J_x & 0 & 0 & 0
\end{array}\right), \\
A^{(\Sigma)}_5 &= i \left(\begin{array}{ccccc}
 0 & -h_z^{(2)} & h_y^{(2)} & 0 & J_x \\
 h_z^{(2)} & 0 & -h_x^{(2)} & -J_y & 0 \\
 -h_y^{(2)} & h_x^{(2)} & 0 & 0 & 0 \\
 0 & J_y & 0 & 0 & -h_z^{(1)} \\
 -J_x & 0 & 0 & h_z^{(1)} & 0
\end{array}\right).
\label{eq:so5AD}
\end{aligned}
\ee
\end{widetext}
 
This is an example case where a D--type DI greatly simplifies the problem. We further observe that there are two blocks in $A^{(\Sigma)}_5$ which are ``coupled'' to each other through $J_x$ and $J_y$.
The first block is a Bloch equation similar to \eq{eq:so4pu1Apm1}, whereas the second block is solved by
\be
g_{X\openone }^2 + g_{Y\openone }^2 = C, \qquad \dot g_{X\openone }^2 + g_{X\openone }^2 = C {h_z^{(1)}}^2.
\label{eq:osciallorlike}
\ee
This feature can be exploited in construction of perturbative solutions in the weak--coupling limit.

\subsection{Solutions of the Bloch Equation}
 We have encountered $A \in \so(3)$ for $\mathfrak I_S \in \so(4) \oplus \mathfrak u (1)$ in Eqs. \eref{eq:so4pu1Apm1}, \eref{eq:so5Apm1} and \eref{eq:so5Apm2} .
Such a matrix lives in the adjoint representation of $\su(2)$, as a result we can find a two--level Hamiltonian satisfying $A^\pm = \ad(H^\pm)$, meaning we have effectively reduced four--level problems with $H \in \so(4)$ and $\so(4)\oplus \mathfrak u(1)$
into a pair of independent two--level problems, associated with the Hamiltonian $H^\pm=(h_x \pm J) \sigma_x + h_y \sigma_y + h_z \sigma_z$ for \eq{eq:so4pu1Apm1}. (see  
Table III).

For the special case of \eq{eq:so4pu1Apm1} with $h_x=\mathrm{const.}$, $h_y = B \cos \omega t$, $h_z = B \sin \omega t$ where $B$ and $\omega$  are time--independent, the Hamiltonian \eq{eq:HIsing} becomes the NMR Hamiltonian in rotating--frame, and the problem has a simple exact solution $\vect g^\pm = (\pm J + h_x -\omega/2,  B \cos \omega t,  B \sin \omega t)^T$ such that we obtain
\be
\begin{aligned}
I_S^\pm =& (\pm J + [h_x-\omega/2])([\openone \pm X]X) +  \\
& B \cos\omega t ([\openone \pm X]Y) + B \sin\omega t ([\openone \pm X]Z).
\end{aligned}
\ee
We note that the Hamiltonian and the invariant are both periodic with $T=2\pi/\omega$ and thus can be used to obtain cyclic non--adiabatic geometric phases \cite{Monteoliva1994,Mostafazadeh1998}. The  unnormalized eigenvectors and eigenvalues of a simpler DI
\be
\begin{aligned}
I_S =& I_S^+ + I_S^- = 2J XX + (2h_x - \omega) \openone X  + \\ 
& 2B\cos \omega t \openone Y + 2B\sin \omega t \openone Z
\end{aligned}
\ee
follow as
\be
\begin{aligned}
\ket{\phi_+^\pm} &= i(g_2^+ + i g_1^+) \left(\begin{array}{c}1\\0\\1\\0\end{array}\right) + (g_3^+ \pm g^+) \left(\begin{array}{c}0\\1\\0\\1\end{array}\right),  \\
\lambda_+^\pm &= \pm g^+;  \\
\ket{\phi_-^\pm} &= i(g_2^- + i g_1^-)\left(\begin{array}{c}-1\\0\\1\\0\end{array}\right) + (g_3^-  \pm g^-)\left(\begin{array}{c}0\\-1\\0\\1\end{array}\right),  \\
\lambda_-^\pm &= \pm g^-
\end{aligned}
\ee
where $g^\pm = |\vect g^\pm|$. Using these eigenvectors in \eq{eq:U}, we can obtain the analytic time--evolution operator of the system. It would be straightforward to design quantum control of such systems accordingly.

Finally, the type of Bloch equations that appeared in \eq{eq:so5Apm1} and \eq{eq:so5Apm2} has several exact solutions \cite{Bagrov2001} corresponding to experimentally realizable fields. When the time--dependent term is periodic and is much greater that the time--independent part (weak--coupling limit), a perturbative solution can be constructed \cite{Barata2002}. 

\begin{table}
\begin{ruledtabular}
\begin{tabular}{l l}
$\ad(\sigma_x) = 2 i \left( \begin{array}{ccc}
0 & 0 & 0 \\
0 & 0 & -1 \\
0 & 1 & 0 \\
\end{array}
\right)$ &
$\ad(\sigma_y) = 2 i \left( \begin{array}{ccc}
0 & 0 & 1 \\
0 & 0 & 0 \\
-1 & 0 & 0 \\
\end{array}
\right)$ \\
$\ad(\sigma_z) = 2 i \left( \begin{array}{ccc}
0 & -1 & 0 \\
1 & 0 & 0 \\
0 & 0 & 0 \\
\end{array}
\right)$ \\
\end{tabular}
\label{tab:su2adjoint}
\end{ruledtabular}
\caption{Adjoint representation of $\su(2)$, $[\ad(\sigma_i)]_{jk}=-i (f_i)_{jk}$ where $[\sigma_i,\sigma_j] = i f_{ijk} \sigma_k$.}
\end{table}

 \section{Discussion and Outlook}
In this work we offered a Lie--algebraic classification and construction of DIs of four--level systems. Such systems cover two--qubit systems which are relevant and essential for quantum computation. Applications of DIs go beyond the exact solutions of the time--dependent Schr\"odinger equation, such as fast control of four--level systems beyond adiabatic regimes and non--abelian geometric phases.
We have shown that the $\mathrm{span}(\mathfrak I) = \so(4)$ and $\mathrm{span}(\mathfrak I) = \so(4)\oplus \mathfrak{u}(1)$ cases are reduced to the solutions of two--level systems, with analytic solutions corresponding to practically realizable four--level parameters. One such solution was explicitly given. The $\mathrm{span}(\mathfrak I) = \so(5)$ was noted for its D--type DI, whose solution can be reduced to solving two weakly coupled sets of differential equations.
These DIs can be applied to gain various control of four--level systems. In an ongoing work to be reported elsewhere, we try to devise a CNOT gate as a control passage based on the DIs constructed in this paper.

\begin{acknowledgments}
This work is partially supported by \lq Open Research Center\rq~Project for Private Universities: matching fund subsidy from the Ministry of Education, Culture, Sports, Science and Technology, Japan (MEXT).
MN would like to thank partial supports of Grants--in--Aid for Scientific Research from the JSPS (Grant No.~24320008). UG is supported by the MEXT scholarship for foreign students. YW thanks Ling--Yan Hung and Robert Mann for helpful discussions.
\end{acknowledgments}

\appendix

\section{No $\su(3)$ S--Type DIs}
\label{sec:nosu3}

\begin{widetext}
\begin{table*}
\begin{ruledtabular}
\begin{tabular}{r r r r r}
$\lambda_1 =
\begin{pmatrix}
0 & 1 & 0 & 0 \\   
1 & 0 & 0 & 0 \\
0 & 0 & 0 & 0 \\
0 & 0 & 0 & 0 \\
 \end{pmatrix}$
&
$\lambda_2 = \begin{pmatrix}
0 & -i & 0 & 0 \\   
i & 0 & 0 & 0 \\
0 & 0 & 0 & 0 \\
0 & 0 & 0 & 0 \\
 \end{pmatrix}$
&
$\lambda_3 = \begin{pmatrix}
1 & 0 & 0 & 0 \\   
0 & -1 & 0 & 0 \\
0 & 0 & 0 & 0 \\
0 & 0 & 0 & 0 \\
 \end{pmatrix}$
&
$\lambda_4 = \begin{pmatrix}
0 & 0 & 1 & 0 \\   
0 & 0 & 0 & 0 \\
1 & 0 & 0 & 0 \\
0 & 0 & 0 & 0 \\
 \end{pmatrix}$
&
$\lambda_5 = \begin{pmatrix}
0 & 0 & -i & 0 \\   
0 & 0 & 0 & 0 \\
i & 0 & 0 & 0 \\
0 & 0 & 0 & 0 \\
 \end{pmatrix}$
\\
$\lambda_6 = \begin{pmatrix}
0 & 0 & 0 & 0 \\   
0 & 0 & 1 & 0 \\
0 & 1 & 0 & 0 \\
0 & 0 & 0 & 0 \\
 \end{pmatrix}$
&
$\lambda_7 = \begin{pmatrix}
0 & 0 & 0 & 0 \\   
0 & 0 & -i & 0 \\
0 & i & 0 & 0 \\
0 & 0 & 0 & 0 \\
 \end{pmatrix}$
&
$\lambda_8 = \frac{1}{\sqrt 3}\begin{pmatrix}
1 & 0 & 0 & 0 \\   
0 & 1 & 0 & 0 \\
0 & 0 & -2 & 0 \\
0 & 0 & 0 & 0 \\
 \end{pmatrix}$
&
$\lambda_9 = \begin{pmatrix}
0 & 0 & 0 & 1 \\   
0 & 0 & 0 & 0 \\
0 & 0 & 0 & 0 \\
1 & 0 & 0 & 0 \\
 \end{pmatrix}$
&
$\lambda_{10} = \begin{pmatrix}
0 & 0 & 0 & -i \\   
0 & 0 & 0 & 0 \\
0 & 0 & 0 & 0 \\
i & 0 & 0 & 0 \\
 \end{pmatrix}$
\\
$\lambda_{11} = \begin{pmatrix}
0 & 0 & 0 & 0 \\   
0 & 0 & 0 & 1 \\
0 & 0 & 0 & 0 \\
0 & 1 & 0 & 0 \\
 \end{pmatrix}$
&
$\lambda_{12} = \begin{pmatrix}
0 & 0 & 0 & 0 \\   
0 & 0 & 0 & -i \\
0 & 0 & 0 & 0 \\
0 & i & 0 & 0 \\
 \end{pmatrix}$
&
$\lambda_{13} = \begin{pmatrix}
0 & 0 & 0 & 0 \\   
0 & 0 & 0 & 0 \\
0 & 0 & 0 & 1 \\
0 & 0 & 1 & 0 \\
 \end{pmatrix}$
&
$\lambda_{14} = \begin{pmatrix}
0 & 0 & 0 & 0 \\   
0 & 0 & 0 & 0 \\
0 & 0 & 0 & -i \\
0 & 0 & i & 0 \\
 \end{pmatrix}$
&
$\lambda_{15} = \frac{1}{\sqrt 6}\begin{pmatrix}
1 & 0 & 0 & 0 \\   
0 & 1 & 0 & 0 \\
0 & 0 & 1 & 0 \\
0 & 0 & 0 & -3 \\
 \end{pmatrix}$
\end{tabular}
\end{ruledtabular}
\caption{$\lambda$ matrices, obeying $\tr(\lambda_i \lambda_j) = 2\delta_{ij}$. $\lambda_{1-3}$ is an $\su(2)$ and $\lambda_{1-8}$ is an $\su(3)$. $\lambda_8$ commutes with $\lambda_{1-3}$, $\lambda_{15}$ commutes with $\lambda_{1-8}$, giving rise to $\su(2) \oplus \mathfrak u(1)$ and $\su(3) \oplus \mathfrak u(1)$ subalgebras.}
\label{tab:lambdas}
\end{table*}
\end{widetext}

\begin{table}
\begin{ruledtabular}
\begin{tabular}{l l l r | l l l r | l l l r}
$i$ & $j$ & $k$ & $f_{ijk}$ & $i$ & $j$ & $k$ & $f_{ijk}$ & $i$ & $j$ & $k$ & $f_{ijk}$\\
\hline
1 & 2 & 3 & 2      &  1 & 9 & 12 & 1         &    6 & 12 & 13 & -1 \\
1 & 4 & 7 & 1      &  1 & 10 & 11 & $-1$     &    7 & 11 & 13 & 1 \\
1 & 5 & 6 & $-1$   &  2 & 9 & 11 & 1         &    7 & 12 & 14 & 1 \\
2 & 4 & 6 & 1      &  2 & 10 & 12 & 1        &    8 & 9 & 10 & $1/\sqrt{3}$ \\
2 & 5 & 7 & 1      &  3 & 9 & 10 & 1         &    8 & 11 & 12 & $1/\sqrt{3}$ \\
3 & 4 & 5 & 1      &  3 & 11 & 12 & $-1$     &    8 & 13 & 14 & $-2/\sqrt{3}$ \\
3 & 6 & 7 & $-1$   &  4 & 9 & 14 & 1         &    9 & 10 & 15 & $2\sqrt{2/3}$ \\
4 & 5 & 8 & $\sqrt 3$ & 4 & 10 & 13 & $-1$   &    11 & 12 & 15 & $2\sqrt{2/3}$ \\
6 & 7 & 8 & $\sqrt 3$ & 5 & 9 & 13 & 1       &    13 & 14 & 15 & $2\sqrt{2/3}$ \\
  &   &   &        &  5 & 10 & 14 & 1        &    6 & 11 & 14 & 1 \\
\end{tabular}
 \end{ruledtabular}
 \caption{Structure constants for $\lambda$ matrices.}
\label{tab:flambda}
\end{table}

The fundamental representation of $\su(4)$ is $15$ Hermitian set of $\{\lambda_i\}$ matrices (Tables \ref{tab:lambdas} and \ref{tab:flambda}). The first $8$ of these, along with $\lambda_{15}$ which commutes with the rest form the $\su(3) \oplus \mathfrak{u}(1)$ subalgebra.
They are related to spinor generators as
\be
2 \lambda_1 = (\openone + Z)  X, \quad & 2 \lambda_2 = (\openone + Z)   Y, \nonumber \\
2 \lambda_3 = (\openone + Z)   Z, \quad & 2 \lambda_4 = X  (\openone + Z), \nonumber \\
2 \lambda_5 = Y  (\openone + Z), \quad & 2 \lambda_6 = X  X + Y  Y, \nonumber \\
2 \lambda_7 = Y  X - X  Y, \quad & 2\sqrt{3} \lambda_8 = 2 Z  \openone - \openone  Z + Z Z, \nonumber \\
\sqrt{6} \lambda_{15} = Z  \openone + \openone  Z - Z  Z. & &
\ee
$2H$ written in terms of $\lambda_j$ is
\be
& & h_x^{(2)} \lambda_1 + h_y^{(2)} \lambda_2 + (h_z^{(2)} + J_z) \lambda_3 + \nonumber \\
& &  h_x^{(1)} \lambda_4 + h_y^{(1)} \lambda_5 + (J_x + J_y) \lambda_6 +   \frac{2 h_z^{(1)} - h_z^{(2)} + J_z}{\sqrt{3}} \lambda_8 +  \nonumber \\
& &  (J_x-J_y) \lambda_9 + h_x^{(1)} \lambda_{11} + h_y^{(1)} \lambda_{12} + h_x^{(2)} \lambda_{13} + h_y^{(2)} \lambda_{14} + \nonumber \\
& &  \frac{2}{\sqrt{3}} (h_z^{(1)} + h_z^{(2)} - J_z) \lambda_{15}
\ee
We see that in order to get rid of $\lambda_{9-14}$ terms, we must have $h_x^{(1)} = h_y^{(1)} = h_x^{(2)} = h_y^{(2)}=0$ and $J_x = J_y$. This, however, reduces the Hamiltonian to
\be
& & (h_z^{(2)} + J_z) \lambda_3   +    2 J \lambda_6   +  \frac{2 h_z^{(1)} - h_z^{(2)} + J_z}{\sqrt{3}} \lambda_8   + \nonumber \\
& &    \frac{2}{\sqrt{3}} (h_z^{(1)} + h_z^{(2)} - J_z) \lambda_{15},
\ee
where we defined $J=J_x=J_y$. The coefficients are linearly independent but the set of generators belong to $\su(2) \oplus \mathfrak{u}(1) \subset \su(3) \oplus \mathfrak{u}(1)$, from which we conclude that one cannot emulate and gain the full SU(3) control of a qutrit using two qubits with a Hamiltonian of the form \eq{eq:H}.

\bibliographystyle{apsrev}
\bibliography{LRI,extra}

\end{document}